# AUTOSTABILISATION DE LA PHASE ACOUSTIQUE DANS UN MIROIR BRILLOUIN INVERSÉ EN PRÉSENCE D'UNE FORTE DISPERSION ACOUSTIQUE

Carlos Montes[1], Éric Picholle[1], Antonio Montes[2]

[1] INPHYNI, Institut de Physique de Nice, UMR 7010, CNRS & Université Côte d'Azur, 06100 Nice

[2] Mathématiques Appliquées II (MA2), Universitat Politècnica de Catalunya, Barcelone, Espagne

eric.picholle@inphyni.cnrs.fr

**RÉSUMÉ**

On considère un régime contre-intuitif de diffusion Brillouin stimulée où la pompe absorbe les défauts de phase de l'onde Brillouin, ce qui induit une auto-stabilisation non-linéaire en phase et un étrécissement spectral de l'onde acoustique.

**MOTS-CLEFS :** *Diffusion Brillouin stimulée ; Dispersion acoustique ; Cohérence acoustique*

## 1. INTRODUCTION

La diffusion Brillouin stimulée (DBS) est un processus d'interaction à trois ondes — deux ondes optiques, "pompe" et "Brillouin" rétrodiffusée, et une onde acoustique. Sa dynamique non linéaire extrêmement riche (régimes périodiques, quasi-périodiques, solitoniques, etc.) est désormais très bien connue dans les fibres optiques, et très bien décrite par un modèle cohérent unidimensionnel à trois ondes prenant en compte un amortissement fini des ondes acoustiques, ainsi que l'effet Kerr optique dans une approximation d'enveloppes lentement variables [1,2]. Le développement d'expériences de DBS dans les fibres microstructurées [3] nous a amenés à y intégrer une description plus sophistiquée de l'évolution de l'onde acoustique, prenant en compte une dispersion acoustique significative [4].

## 2. MODÈLE INERTIEL ET MIROIR BRILLOUIN INVERSÉ

En présence d'une forte dispersion acoustique, si l'évolution des amplitudes complexes des enveloppes des ondes pompe et Brillouin, $E_p$ & $E_B$ reste bien décrites par le système habituel :

$$(\partial_t + c/n\partial_x + \gamma_e)E_p = -K_{SBS}E_B E_a, \qquad (1)$$

$$(\partial_t - c/n\partial_x + \gamma_e)E_B = K_{SBS}E_p E_a^*, \qquad (2)$$

il convient de prendre en compte des termes d'ordre supérieur dans la description de celle de l'enveloppe acoustique, $E_a$, pour aboutir au modèle inertiel de la DBS (Éqs. 1, 2 & 3). On a alors :

$$[(1+2i\alpha)\partial_t + c_a\partial_x + i\alpha(\partial_{tt} - \varepsilon^2\partial_{xx} + \gamma_a]E_a = K_{SBS}E_p E_B^* \qquad (3)$$

où $\alpha = c_a \frac{d^2 k_a}{d\omega_a^2}(\omega_a) K_{SBS}\sqrt{I_P^0}/2\omega_a$ est le paramètre de dispersion acoustique et $\varepsilon = c_a/c$ ~$10^{-5}$ le rapport des vitesses acoustique et optique.

Une conséquence inattendue de cet approfondissement est la mise en évidence de la possibilité, pour de fortes dispersions acoustiques, de miroirs Brillouin inversés, où l'amplitude acoustique *croît* avec la distance dans des résonateurs Brillouin (Fig. 1 a&b, partie supérieure).

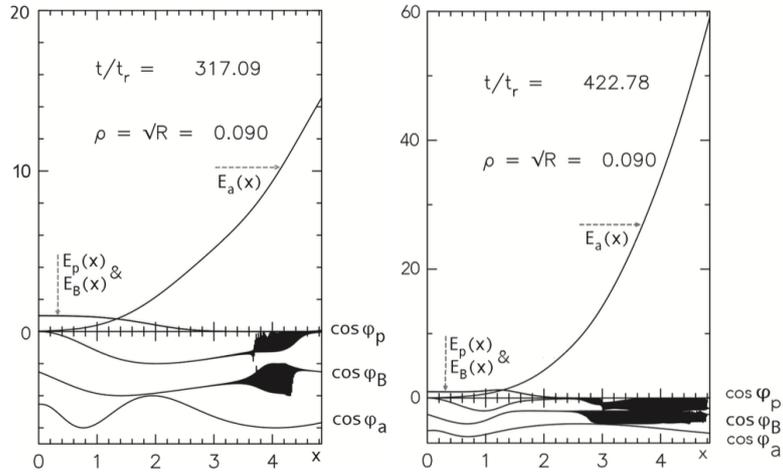

Fig. 1 : Miroirs Brillouin présentant une réflexion totale avec des profils acoustiques inversés (régimes asymptotiques pour les intensités). (a) à g.: dispersion modérée : $\alpha = 0,5$, $L/\Lambda = 3,86$, $G = 0,35$, $\mu = 22,07$. (b) à dr. : dispersion forte : $\alpha = 1$, $L/\Lambda = 1,84$, $G = 0,55$, $\mu = 17,60$ [Numérique] [4].

L'accumulation de l'onde acoustique à l'extrémité de la fibre opposée à celle d'injection de la pompe induit une très grande efficacité de la DBS dans cette région, avec pour conséquence, d'une part, d'une très grande efficacité de ces miroirs Brillouin, avec des taux de réflexion proches de 100 %, et d'autre part, en l'absence de réinjection de la pompe, une très faible intensité de celle-ci et une variation très rapide des phases pompe et Brillouin (zones noires non résolues dans la partie inférieure des Fig. 1 a&b).

### 3. DYNAMIQUE DES PHASES

Plus spécifiquement, on observe dans cette région une nette anti-corrélation de l'évolution spatiale des phases des ondes pompe et Brillouin, et une quasi-stationnarité de celle de l'onde acoustique (Fig. 2).

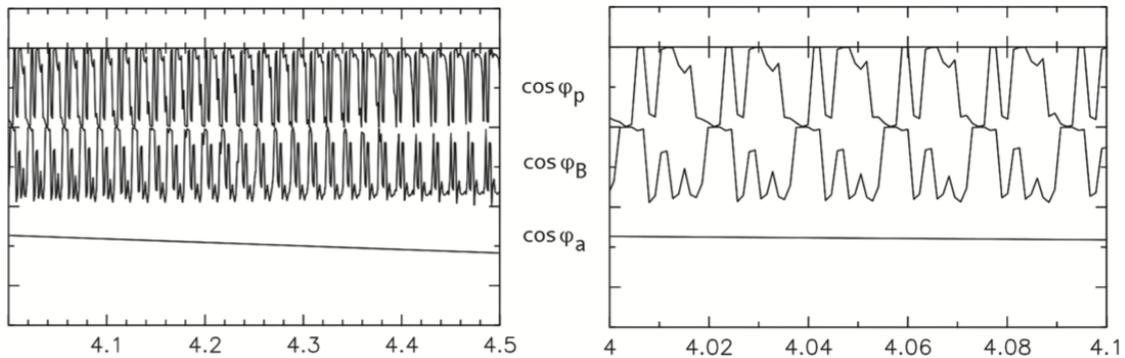

Fig. 2 : Zooms successifs sur l'évolution spatiale des phases de la Fig. 1b.

*Mutatis mutandis,* ce phénomène est réminiscent de celui déjà identifié dans les résonateurs Brillouin classiques, où c'est la phase Brillouin qui est ainsi stabilisée [5], permettant le développement de lasers Brillouin ultracohérents [6].

Très schématiquement, hors résonance, le processus DBS tend à maximiser le gain, lui-même proportionnel à *cos* $\Phi$, où $\Phi = \varphi_p - \varphi_B - \varphi_a$ ; dans la mesure où les phases des trois ondes tournent d'autant plus vite qu'elles sont moins intenses *($\partial \varphi_p \propto 1/E_i$)*. Dans le cas classique, c'est donc l'onde acoustique, de faible amplitude, qui est en permanence éteinte et reconstruite avec la phase pertinente, dans une alternance de processus Brillouin Stokes et anti-Stokes, et "absorbe" les défauts de phase de l'onde de pompe *($\varphi_a(t) = \varphi_p(t) \rightarrow \varphi_B = \Phi = 0$)*. Or dans le cas d'un miroir inversé, c'est au contraire l'amplitude de la pompe qui peut devenir très inférieure à celle des deux autres ondes, et absorber les fluctuations de phase de l'onde Brillouin *($\varphi_p(t) = \varphi_B(t) \rightarrow \varphi_a = \Phi = 0$)*. (Pour simplifier, on a ici raisonné à la résonance exacte ; dans le cas général, légèrement hors résonance, la correction peut inclure un gradient de phase, visible sur la partie gauche de la Fig. 2). Le rapport entre la largeur spectrale de l'onde ainsi stabilisée en phase et celle de l'onde pompe est de l'ordre de $\varepsilon^{-1} = c/c_a \sim 10^5$.

Il apparaît donc que, en régime de miroir Brillouin inversé et en présence d'une forte dispersion acoustique, la diffusion Brillouin stimulée est susceptible de produire des ondes hypersonores (typiquement 10 GHz) d'une très grande pureté spectrale, typiquement de l'ordre du Hz pour une pompe optique MHz.

### Références

[1] J. Botineau, C. Leycuras, É. Picholle et C. Montes, "Stabilization of a stimulated Brillouin fiber ring laser by strong pump modulation", J. Opt. Soc. Am. B, **6**, 3, pp. 300-312 (1989).

[2] R.W. Boyd, K. Rzaewski et P. Narum, "Noise initiation of stimulated Brillouin scattering", Phys. Rev. A, **42**, p. 5514 (1990).

[3] R.I. Woodward, É. Picholle et C. Montes, "Brillouin solitons and enhanced mirror in the presence of acoustic dispersion in a small-core photonic crystal fiber", JNOG'35, Rennes (2015).

[4] A. Montes, C. Montes et É. Picholle, "SBS mirror with inverted acoustic profile in presence of strong acoustic dispersion", J. Opt. Soc. Am. B, **38**, 2, pp. 456-465 (2021).

[5] É. Picholle, C. Montes, C. Leycuras, O. Legrand et J. Botineau, Observation of dissipative superluminal solitons in a Brillouin fiber ring laser, Phys. Rev. Lett., **66**, 1454-1457 (1991).

[6] S. Gundavarapu *et al*., "Sub-Hertz fundamental linewidth photonic integrated Brillouin laser," Nat. Photonics **13**, 60–67 (2019).